\documentclass[12pt]{article}
\def\be{\begin{equation}}
\def\ee{\end{equation}}
\def\ba{\begin{eqnarray}}
\def\ea{\end{eqnarray}}
\def\la{\langle}
\def\ra{\rangle}

\def\m{\mu}
\def\n{\nu}
\def\h{\hskip 1cm}

\def\A1{A_{-1}}
\usepackage{epsfig}
\begin{document}
\begin{titlepage}
\vspace{3cm}
\begin{center}{\Large \bf Two-qubit state sharing between N parties using only Bell pairs and Bell-measurement}\\
\vspace{1cm}\h Razieh Annabestani\footnote{email:
razie$_{-}$anabestani@physics.sharif.ir},\h  Vahid Karimipour
\footnote{ email: vahid@sharif.edu}\\
\vspace{1cm} Department of Physics, Sharif University of Technology,\\
P.O. Box 11155-9161,\\ Tehran, Iran
\end{center}
\vskip 3cm
\begin{abstract} A quantum protocol for sharing an arbitrary two-qubit state
between N parties is introduced. Any of the members, can retrieve
the state, only with collaboration of the other parties. We will
show that in terms of resources, i.e. the number of classical
bits, the number of Bell pairs shared, and also the type of
measurements, our protocol is more efficient. For achieving this,
we introduce the basic technique of secure passing of an unknown
two qubit state among a sequence of parties, none of which can
retrieve the state without authorization of the sender and the
other members of the group.

\end{abstract}
\end{titlepage}

\section{Introduction}\label{intro}

The problem of secure sharing of an unknown quantum state between
two parties \cite{BuzekHillery, Cleve, Li, Lance, Deng, chineese},
is an important one in quantum cryptography, the branch of quantum
information science which uses quantum correlations between distant
parties for establishing cryptographic keys and splitting and
sharing information between different parties ( For recent reviews
see \cite{ReviewGisin, ReviewZhao, Review1} and references therein).
The security of all these protocols are based not on mathematical
theorems, but on the very foundations of quantum mechanical laws,
where any attempt of eavesdropping can be detected
by affecting the quantum correlations between legitimate parties.\\

The first protocol for sharing a quantum state between two parties,
was suggested in \cite{BuzekHillery}, where a GHZ \cite{GHZ} state
of the form
\begin{equation}
\label{ghz}
    |GHZ\ra=\frac{1}{\sqrt{2}}(|000\ra+|111\ra),
\end{equation}
is shared between the sender Alice and the receivers Bob and
Charlie. For sending an unknown qubit state $|q\ra:=a|0\ra+b|1\ra$,
Alice makes a Bell measurement on this and his share of the GHZ
state, obtaining one of the four possible results. She then
announces the result of her measurement in the form of two classical
bits and asks either Bob or Charlie to perform a measurement on
their share in the $X$ basis and announce their results. The other
party who has done no measurement can now retrieve the state by a
suitable unitary operation. The fact that it is Alice who determines
who is to receive the final state and who should do measurement in
the $X$ basis, guarantees the security of the protocol against the
cheating of both the receivers.\\

While there is straightforward way to generalize this protocol for
sharing a qubit to an arbitrary number of receivers, it is far from
obvious how to generalize this protocol in the other direction, that
is, sending a two-qubit state to two or an arbitrary number of
parties. It is true that the original scheme of \cite{BuzekHillery}
can be directly generalized for sharing of arbitrary $d$-level
states, simply by generalizing Bell states and Pauli operators in a
straightforward way \cite{dlevel1} \cite{dlevel2}, however such
generalization require physical implementation of $d-$level states
or qudits, while the mainstream candidates for implementation of
quantum information processing are based on two-level states or
qubits. Therefore it is essential to generalize these and any other
protocol for quantum information to higher dimensional states which
represent not the state of a single quantum entity (like a qudit),
but that of a few
quantum objects (photons, ions,...) or qubits.\\

\begin{figure}[h]
 \centering
  \includegraphics[width=10cm,height=4cm,angle=0]{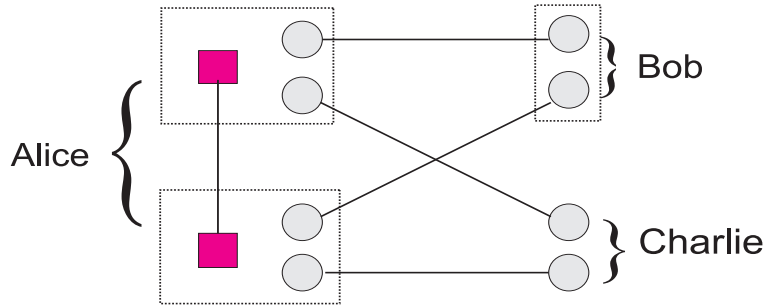}
\caption{\small The pattern of Bell state sharing and measurements
in the protocol of \cite{chineese}. The light gray bulbs indicate
Bell pairs and the red square boxes show the two-qubit state which
is to be shared. The dotted boxes show the GHZ measurements which
Alice has to perform and the two qubit product measurement by
Bob.} \label{chineesefigure}
\end{figure}

It seems that for increasing the number of receivers or the
dimension of the state which is to be sent, there is no other way
other than using multi-party entangled states, like GHZ states and
their generalization. For example in \cite{chineese} a scheme has
been proposed for sharing a two-qubit state between two parties,
where the sender Alice shares two Bell pairs with Bob and Charlie.
The pattern of Bell state sharing and measurements is shown in
figure (\ref{chineesefigure}). To send the two qubit state (shown
with red square boxes) to Charlie, she makes two GHZ measurements
on her shares and the two qubit state which is to be sent and
publicly announces the results of these two measurements. Then Bob
makes a product measurement $\sigma_x\otimes \sigma_x$ on his
qubits, and publicly announces his result. A direct calculation
then shows that Charlie can retrieve the original state after
performing a suitable unitary operation, determined by the results
of public bits announced by Alice and Bob.

\begin{figure}[h]
 \centering
  \includegraphics[width=9cm,height=7cm,angle=0]{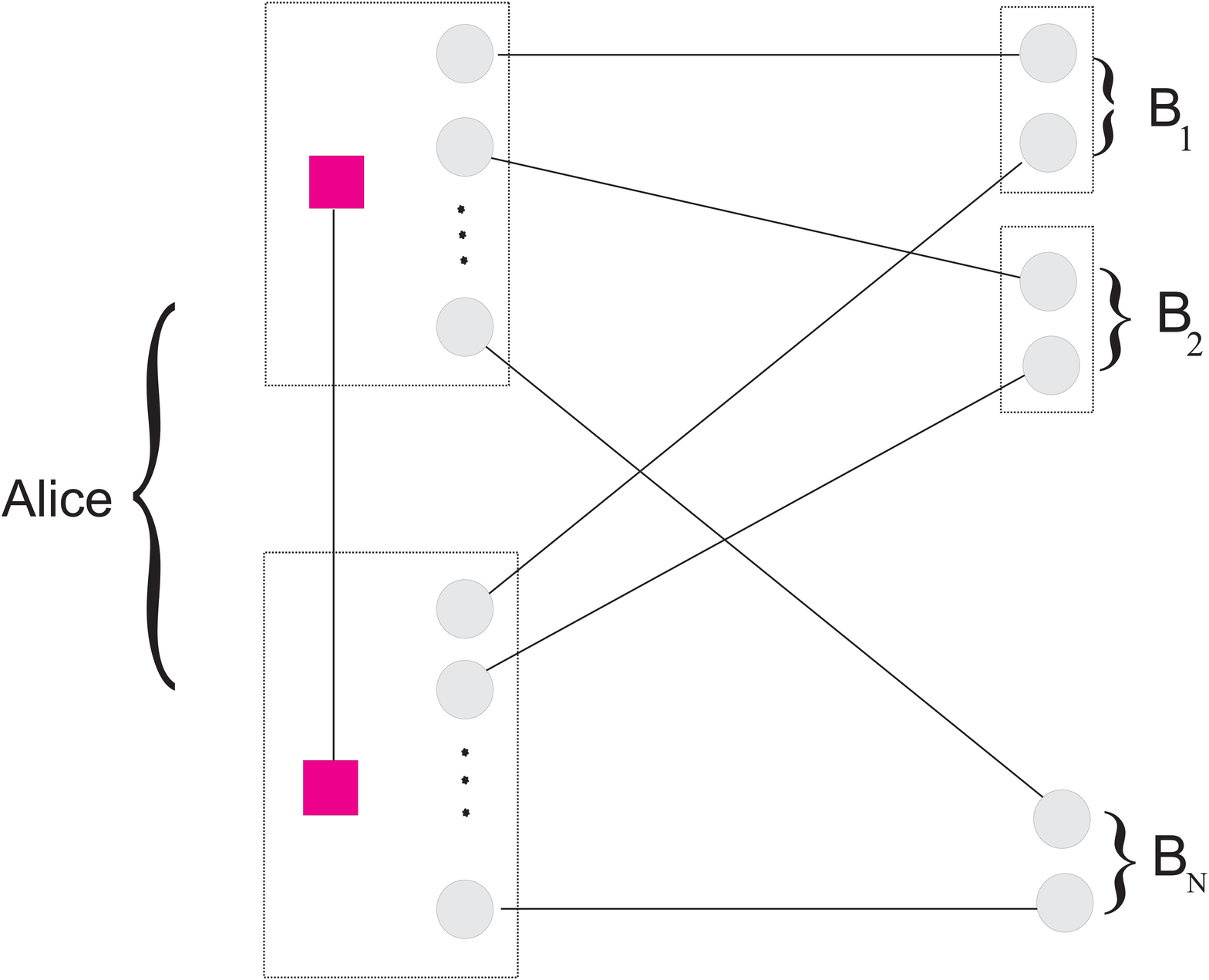}
\caption{\small The pattern of Bell state sharing and measurements
in the protocol of \cite{chineese} for N-parties. The conventions
are the same as in figure (\ref{chineesefigure}). However now
Alice has to perform generalized measurements of GHZ types. }
\label{chineesefigureN}
\end{figure}

Generalization of this protocol for sending a two-qubit state to
$N$ receivers \cite{chineese}, shown in figure
(\ref{chineesefigureN}), requires $2N$ Bell pairs and two
$N-$quibt GHZ measurements, which is a formidable resource,
especially with regard to the GHZ measurements.

In this paper we want to present an alternative scheme for this
kind of state sharing which uses only Bell states and Bell
measurements. The motivation of this work is clear, since
construction, distribution and maintenance of GHZ states and their
generalized N-party states and also performing measurements in
such bases, are much more difficult than that of two party
entangled states. In fact many of the experimental obstacles in
dealing with pairs of entangled states has already been removed
and it is now well known that experimental realization of quantum
key distribution \cite{QKDexp1} \cite{QKDexp2}, and quantum
teleportation \cite{QTELexp} has been
achieved over increasingly long distances.\\

We will show that it is possible to securely share a two-qubit state
with an arbitrary number of receivers so that only one of the
members will be able to retrieve the state with the help of the
other members. We will first demonstrate this for two receivers and
then generalize it to $N$ receivers. For this later part, we
introduce a technique which enables the members to pass the
two-qubit state one by one, among themselves, without any one of
them being able to discover the identity of the state. The essence
of our method is a combination of teleportation and the passing
technique mentioned above. In addition to alleviating the need for
using GHZ states, we will also show that our protocol requires much
less classical bits to be announced publicly by the members. \\

For definiteness, we call the final member who is to receive the
state, the receiver and the other members whose collaboration are
necessary for the receiver to reconstruct the state, simply as
controllers. Like the original scheme of Buzek, Berthiaume and
Hillery, the security of the protocol against cheating of any
subgroup of the members is guaranteed by the fact that it is
Alice, the sender who decides who
is to be the receiver among the members.\\

A remark is in order with regard to our method of presentation and
reasoning. Usually secret sharing schemes, specially between
N-parties, require writing  many party states with lots of indices
and following how these states change or collapse under different
kinds of measurements which even make the notation and the
presentation even more clumsy. Instead of this, we use a
transparent graphical method whose correspondence with the states
and their change is represented in section ($2$). After that we
will use extensively this correspondence and use diagrams to
present our reasoning and results in sections ($3$) and ($4$).
Also we discuses about cheating and the number of resources in
section ($5$) and ($6$). We conclude the paper with a discussion.

\section{Two-qubit state sharing between two parties}
Usually secret sharing schemes, specially between N-parties,
require writing  many party states with lots of indices and
following how these states change or collapse under different
kinds of measurements. This makes the presentation cumbersome and
difficult to follow.  Instead of this, we use a transparent
graphical method whose correspondence with the states and
transformations are explained in the following.
\subsection{Graphical notation}
\begin{figure}[h]
 \centering
  \includegraphics[width=12cm,height=3.5cm,angle=0]{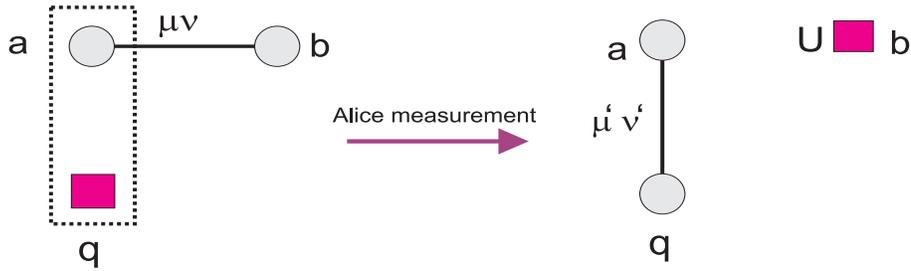}
\caption{\small Teleportation of an unknown qubit $\chi$. The
reterival operator $U=Z^{\mu+\mu'}X^{\nu+\nu'}$ depends on the
initial Bell pair $\mu\nu$ and the result $\mu'\nu'$ of Alice's
measurement, represented by the dashed box.} \label{Teleportation}
\end{figure}

In this section, we briefly remind the teleportation scheme and
set up our graphical conventions. The four Bell pairs are denoted
by $\{\phi_{\mu,\nu}, \mu,\nu=0,1\}$ where
\begin{equation}
|\phi_{\mu,\nu}\ra:=\frac{1}{\sqrt{2}}\sum_{k}(-1)^{\mu
k}|k,k+\nu\ra),
\end{equation}
inversion of which yields
\begin{equation}\label{mn}
|m,n\ra:=\frac{1}{\sqrt{2}}\sum_{\mu}(-1)^{\mu
m}|\phi_{\m,m+n}\ra.
\end{equation}
Let Alice and Bob share a Bell pair of the form $|\phi_{\m,\n}\ra$
and suppose that Alice wants to teleport a qubit (q) in an unknown
state $|\chi\ra=\sum_m a_m |m\ra$ to Bob. We designate the qubits
in possession of Alice and Bob respectively by the subscripts $a$
and $b$. In figure (\ref{Teleportation}), the Bell pair is shown
the two gray-colored bulbs joined by a line with a label $\mu,\nu$
and the state is shown by a red-colored square box. The total
state of (qab) will be in the form
\begin{equation}\label{qab}
    |\Psi\ra_{qab}=|\chi\ra_q|\phi_{\m,\n}\ra_{ab}=\frac{1}{\sqrt{2}}\sum_{m,k}(-1)^{\mu k}a_m
    |m,k\ra_{qa}|k+\n\ra_b.
\end{equation}
Using (\ref{mn}) and rearranging terms, we find that this state
can be rewritten as
\begin{equation}\label{qab}
    |\Psi\ra_{qab}=\frac{1}{2}\sum_{\m',\n'}(-1)^{\mu' \n'}|\phi_{\m',\n'}\ra_{qa}
    (X^{\n+\n'}Z^{\m+\m'}|\chi\ra_b).
\end{equation}
Therefore, the Bell measurement of Alice, performed on the qubits
$a$ and $q$, will project these two qubits onto one of the Bell
pairs $\phi_{\mu'\nu'}$ and the qubit $b$ of Bob to the state
$U|\chi\ra$, where
 $U=Z^{\m+\m'}X^{\n+\n'}$ and $X$ and $Z$ are Pauli
matrices, $Z=|0\ra\la 0|-|1\ra\la 1|$ and $X=|0\ra\la 1|+|1\ra\la 0|$.\\

Upon public announcement of the result of Alice's measurement,
that is the pair of indices $(\mu',\nu')$, Bob can recover the
state $|\chi\ra$ by the action of the operator $U$. This process
is shown in figure (\ref{Teleportation}), where a Bell pair
$|\phi_{\m\n}\ra$ is shown by two gray circles joined by a line on
it with the label $\m\n$ and the unknown state $|\chi\ra$ with a
small red square. The rectangular dotted box, indicates the
measurement of Alice. We will use these conventions in all the
subsequent discussions and diagram. It is important to note that
even if the qubit $q$ is in a mixed state, i.e. if it is part of
larger pure state, the above diagram is still true. To see this
suppose that the qubit $q$ is a part of a two partite system $qq'$
whose state can be written as
\begin{equation}\label{ensemble}
    |\Phi\ra =\sum_i |\chi_i\ra\otimes | \phi_i\ra.
\end{equation}
Since the teleportation scheme in figure (\ref{Teleportation})
does not depend on the state which is to be sent, this means that
the outcome of the protocol will be given by
\begin{equation}\label{ensemble}
    |\Psi\ra =\sum_i (U|\chi_i\ra)\otimes | \phi_i\ra = (U\otimes I)|\Phi\ra.
\end{equation}
Therefore this basic diagram can be used not only when the qubit
is in a pure states, but also when it is part of a larger
multipartite state. When working with two-qubit states which
generically are not symmetric, we denote them graphically by an
arrow, where the arrow goes from the first qubit to the second
qubit.  We will use this fact extensively in the following
discussion and diagrams.

\subsection{Sharing of a two-qubit states between two parties}
\begin{figure}[t]
 \centering
  \includegraphics[width=15cm,height=7cm,angle=0]{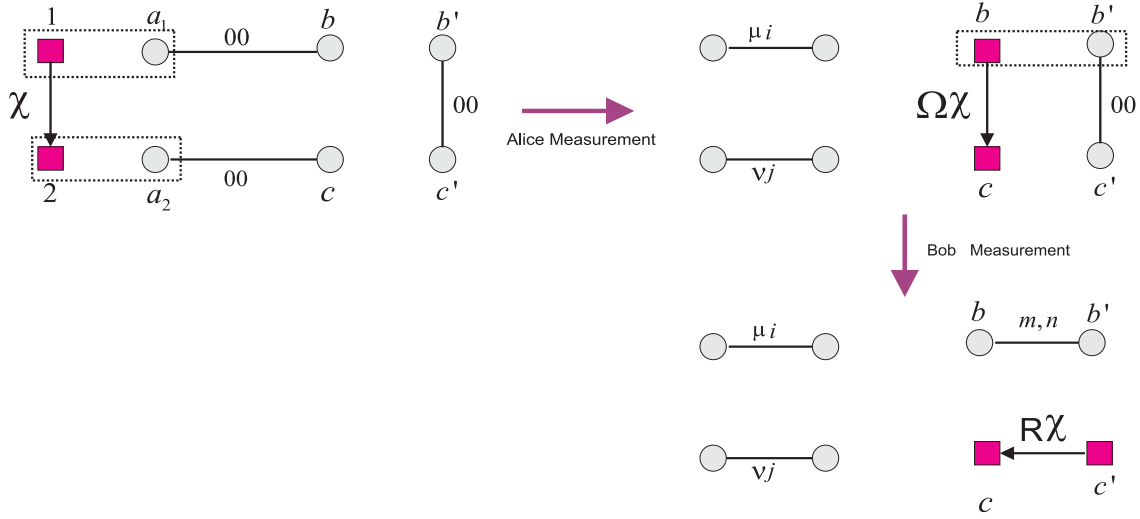}
 \caption{\small Sharing of a two qubit state between Bob and Charlie by Alice.
  Alice holds the qubits $(a_1,a_2)$, Bob holds $(b,b')$ and Charlie hods $(c,c').$ After measurements by Alice and Bob, the states is transferred to Charlie, who can
  recover it by the operator $R$, only with the collaboration of Alice and Bob.} \label{TwoQubitSharing}
\end{figure}

We can use the teleportation scheme of the last section to show
that Alice can share a two-qubit state
 \be\label{chi}
|\chi\ra_{1,2}=\sum_{i,j}K_{ij}|ij\ra, \ee with Bob and Charlie so
that only one of them can retrieve the information by the
collaboration of the other. The pattern of Bell pair sharing and
order of measurements are shown in figure \ref{TwoQubitSharing}.
Alice shares one pair $(a_{1},b)$ with Bob and another identical
$(a_{2}, c)$ whit Charlie, who in turn share a pair $(b',c')$
between themselves. Without loss of generality, we assume all the
EPR pairs to be of the form
$|\phi_{00}\ra:=\frac{1}{\sqrt{2}}|00\ra=|11\ra$. Alice makes two
Bell measurements on the qubits in her possession, namely
$(1,a_1)$ and $(2,a_2)$. In this way he projects these two qubits
onto two Bell states, say $\phi_{\mu,i}$ and $\phi_{\nu,j}$
respectively. From Eq. (\ref{qab}), the state $|\chi\ra$ modified
by the operator $$\Omega= Z^{\mu}X^{i}\otimes Z^{\nu}X^{j}$$ is
transferred to the qubits $b$ and $c$ in possession of Bob and
Charlie. This is shown in the first part of figure
(\ref{TwoQubitSharing}). At this stage the retrieval operator
$\Omega$ is not known to any of them, since Alice can defer her
public announcement of the result to a later stage. Even if she
announces her results, neither Bob nor Charlie, can retrieve the
whole state, since none of them has access to both the qubits.
In case that Alice wants Charlie to be the final owner of the
two-qubit state, she asks Bob to perform a measurement and publicly
announces his result. The two qubits will now be transferred to
Charlie who after Alice's public announcement of her results
$(\m,i)$ and $(\n,j)$, can completely recover the state $|\chi\ra$
by the action of the operator
\begin{equation}\label{R}
R:=(I\otimes Z^{m}X^{n})(Z^{\m}X^{i}\otimes Z^{\n}X^{j})_{cc'}.
\end{equation}
Note that at the end the two Bell pairs will remain between Alice,
Bob and Charlie who can use them for the next round.

\begin{figure}[t]
 \centering
  \includegraphics[width=9cm,height=3cm,angle=0]{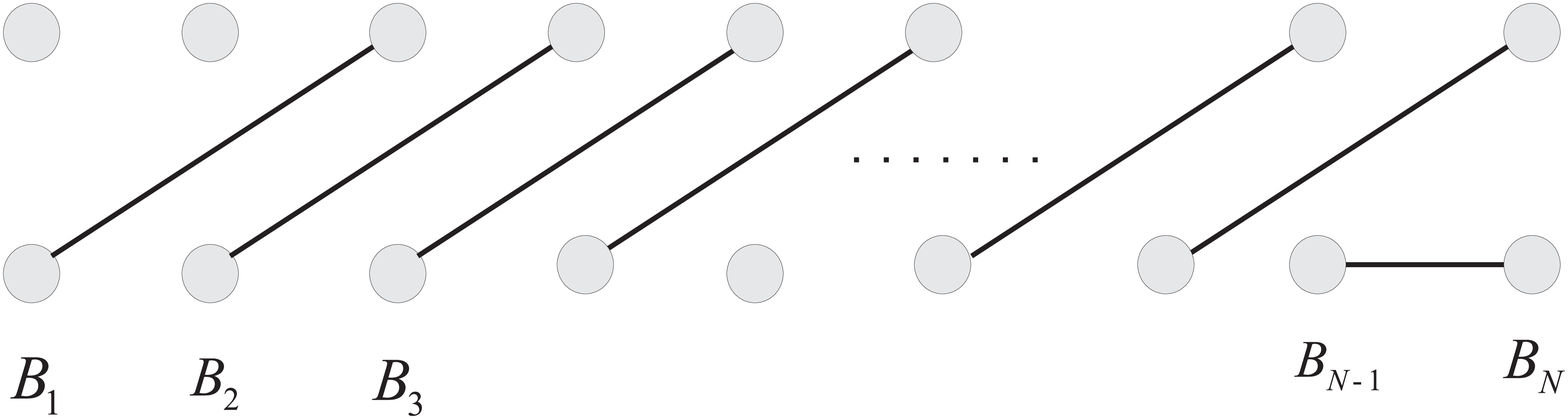}
 \caption{\small The pattern of Bell pairs shared between  N parties $B_{1}, B_{2},...., B_{N}$. Except the first two ($B_1, B_2$) and the
 last two  ($B_{N-1}, B_{N}$) any member $B_{i}$
 shares two Bell pairs with his or her next-nearest neighbors. No two members of the group share two complete Bell pairs. Note that all the Bell pairs are of
 the form $\phi_{00}$. We have not written the labels $00$ on them. } \label{BellPairPattern}
\end{figure}
\section{Two-qubit state sharing between N parties}
The method presented above, for splitting a two-qubit state and
sending it to two parties, can be extended to the general N-party
case. The essential point is to distribute the Bell pairs between
the parties according to a special pattern, shown in figure
(\ref{BellPairPattern}).  Each of the of members except the first
two, holds one share of two Bell pairs, but with two different
members. No two members of the group share two complete Bell pairs.
For example the member $B_3$ shares a Bell pair with $B_1$ and
another one with $B_5$. This scheme of Bell pair sharing, as we will
show, allows all the members to pass a two-qubit state one by one
from the beginning of the chain to the end, without any member being
able to decipher the identity of the state. Here we assume that the
final member of the group, $B_N$ is the one who is going to recover
the state, once the other members of the group $B_1, B_2, \cdots
B_{N-1}$, i.e. the controllers, collaborate with him or her by
publicly announcing the results of their measurements. This explains
the difference of Bell-sharing of him or her with the other members.
Also since Alice feeds the unknown state to the left hand side of
the chain, that is to $B_1$ and $B_2$, they have a different scheme
of Bell pair sharing. In later subsections we will discuss how Alice
can demand that a different member acts as the receiver and hence
secure the protocol against cheating. But before going into these
issues, let
us first describe the protocol itself which runs as follows.\\

\begin{figure}[h]
 \centering
\includegraphics[width=12cm,height=6cm,angle=0]{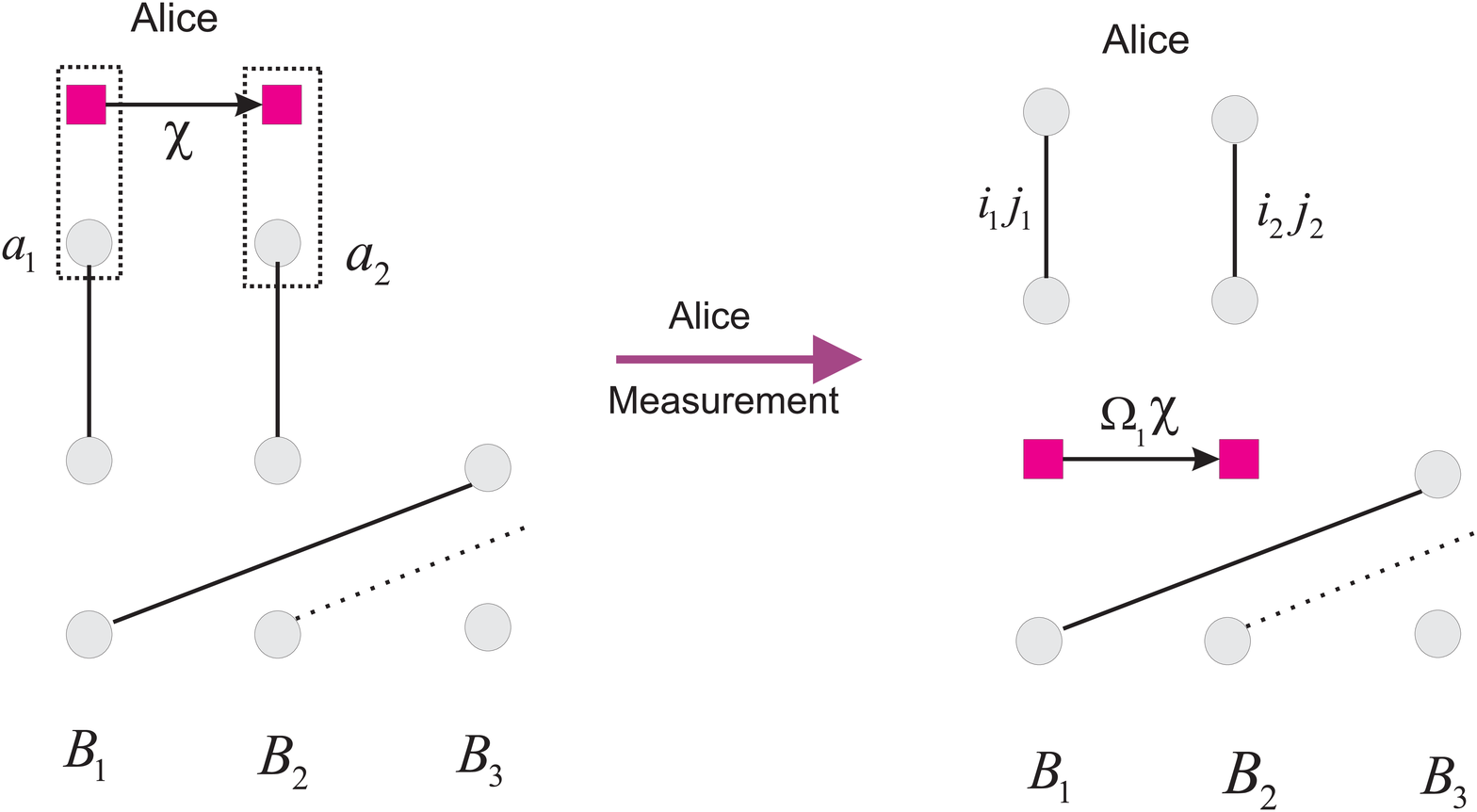}
 \caption{Step one: Alice feeds the state $|\chi\ra_{1,2}$ to the controllers $B_{1}$ and $B_{2}$ who will pass
  their share to their next nearest neighbors one by one to the end of the chain. At this stage the state $|\chi\ra$ can be recovered by
  the action of the operator $\Omega_1=Z^{i_1}X^{j_1}\otimes Z^{i_2}X^{j_2}$.}
\label{Feeding}
\end{figure}

\begin{figure}[b]
 \centering
  \includegraphics[width=15cm,height=3cm,angle=0]{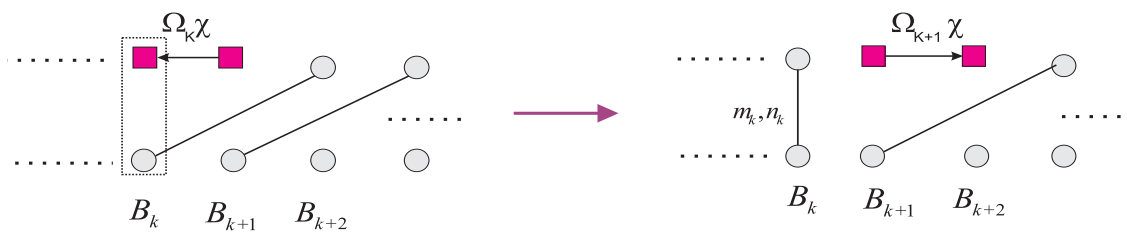}
 \caption{Step two: Passing of the state along the chain. The controllers pass the state one to the other.
 Here the controllers $(B_{k}, B_{k+1})$ pass the state to $(B_{k+1},B_{k+2})$. For this $B_{k}$
 makes a Bell-measurement, shown with the dotted box, with the random result $m_{k}, n_{k}$, leaving his two qubits in a state
 $\phi_{m_{k},n_{k}}$, and teleporting his share to $B_{k+2}$.} \label{Passing}
\end{figure}

\begin{figure}[h]
 \centering
\includegraphics[width=15cm,height=3.7cm,angle=0]{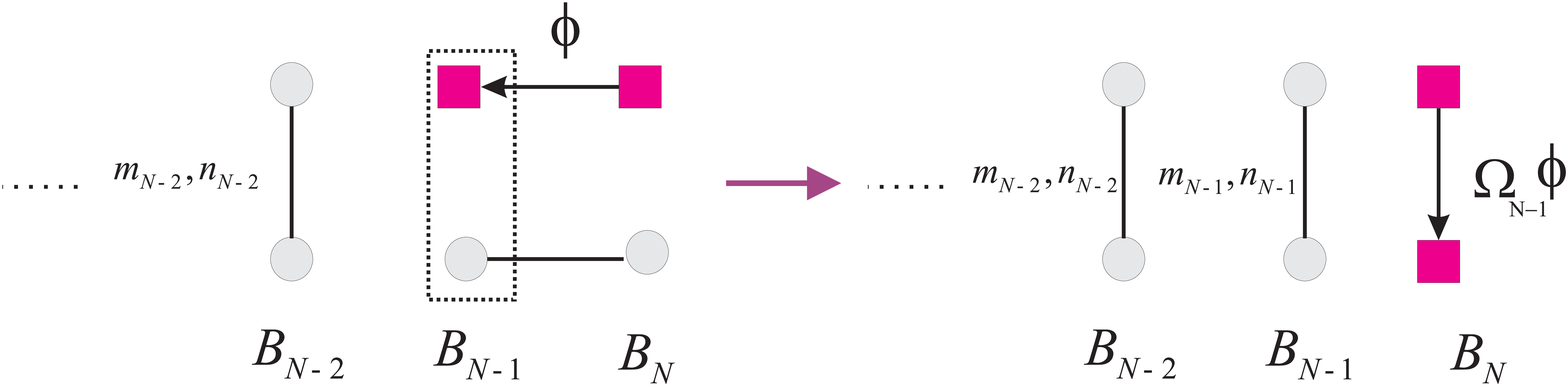}
 \caption{\small Step three, the final step of the protocol. The final controller passes the unknown state $\chi$ to the receiver
 who retrieve it with the operator $R$, determined by the code announced by Alice and all the controllers.}
\label{Retrieval}
\end{figure}

The scheme consists of three major steps, which we may call i)
feeding the state to the chain by Alice, ii) passing the state by
controllers, and iii) retrieval of the state by the receiver.
These three steps are shown respectively in figures
(\ref{Feeding}), (\ref{Passing}) and (\ref{Retrieval}).

Alice first follows the steps of the previous section and
teleports the state $|\chi\ra$ to the left hand side of the chain,
that is to $B_1$ and $B_2$. This is shown in figure
(\ref{Feeding}). Once fed into the chain, the controller $B_1$ can
proceed along the same procedure and teleport his share of the
state to $B_3$. Now the state is shared between $B_2$ and $B_3$.
Upon continuing this process, the state is transferred  one by one
along the chain, until at the end it reaches the controller
$B_{N-1}$ and the receiver $B_N$. A typical intermediate step is
shown in figure (\ref{Passing}). In each step note that the basic
rule of transformation of the diagram is the one displayed in
figure
(\ref{Teleportation}). \\

The operator which will recover the original state is
$\Omega_1=Z^{i_1}X^{j_1}\otimes Z^{i_2}X^{j_2}$. In the next step,
the controller $B_2$ will teleport his share to the the next
nearest member $B_4$, which will cause the state to be shared
between $B_3$ and $B_4$. This process continues to the end, until
the state reaches the last controller $B_{N-1}$ and the receiver
$B_N$. At each step say the $k-th$ step, when the controller $B_k$
makes a Bell measurement with the result $(m_k,n_k)$, the state in
transfer gains an extra correction operator $$U_k=(I\otimes
Z^{m_k}X^{n_k}),$$ leading to the recovery operator
$$\Omega_{k+1}=U_{k+1}\Omega_k.$$ Therefore when the state reaches the
last controller $B_{N-2}$ it has gained the unitary operator
$\Omega_{N-2} $ (figure \ref{Retrieval}). His measurement will
pass the state entirely to the receiver $B_N$, figure
(\ref{Retrieval}) with a further correction $\Omega_{N-1}$. The
receiver can now retrieve the state by the action of the
correction operator

\begin{equation}\label{Rfinal}
    R=\Omega_{N-1}=(Z^{i_1}X^{j_1}\otimes
    Z^{i_2}X^{j_2}) (I\otimes Z^{m_1+m_2+\cdots m_{N-1}}X^{n_1+n_2+\cdots
    n_{N-1}}).
\end{equation}

The correction operator depends on the four bits
$(i_1,i_2,j_1,j_2)$ announced by Alice and the $2(N-2)$ bits
$(m_1, n_1; m_2,n_2; \cdots m_{N-1}, n_{N-1}) $ announced by the
controllers.

After this step, if one of them makes a measurement on his own
qubits; he passes his qubit's context to his next neighbor. If all
of the members keep on this algorithm and exchange their
information with each other, figure \ref{Feeding}, and finally the
receiver can regenerate the initial state only with acting an
appropriate operator. However non of the controllers can extract
all the information; because each of them in each step has one
part of the state $\chi$, so their density matrices are in the
mixed state. It is clear that in the last step, the pattern of
passing is the same as two parties scheme completely, figure \ref{TwoQubitSharing}.\\
\\

\section{Changing the receiver}
At first sight it appears that the pattern of Bell pair sharing
shown in figure (\ref{BellPairPattern}) already fixes who is to be
the receiver of the state among the members of the group, and
therefore Alice as the sender of the state has no choice in
demanding that a different member of the group be the final
receiver of the state. However this pattern can easily be changed
to other desirable patterns by simple entanglement swapping.
Therefore if we change the order of measurements of the members of
the group, we can choose anyone of the members to act as the
receiver and the other ones as controllers. This exchange which is
achieved by a sequence of entanglement swapping, can be demanded
before or after step one, namely the feeding of the state into the
chain. The explicit process is that Alice announces the order of
measurements that the members of the group have to perform. In the
process of these measurements, the identity of the members as
controllers or the receiver will be established.

\begin{figure}[h]
 \centering
\includegraphics[width=14cm,height=3.6cm,angle=0]{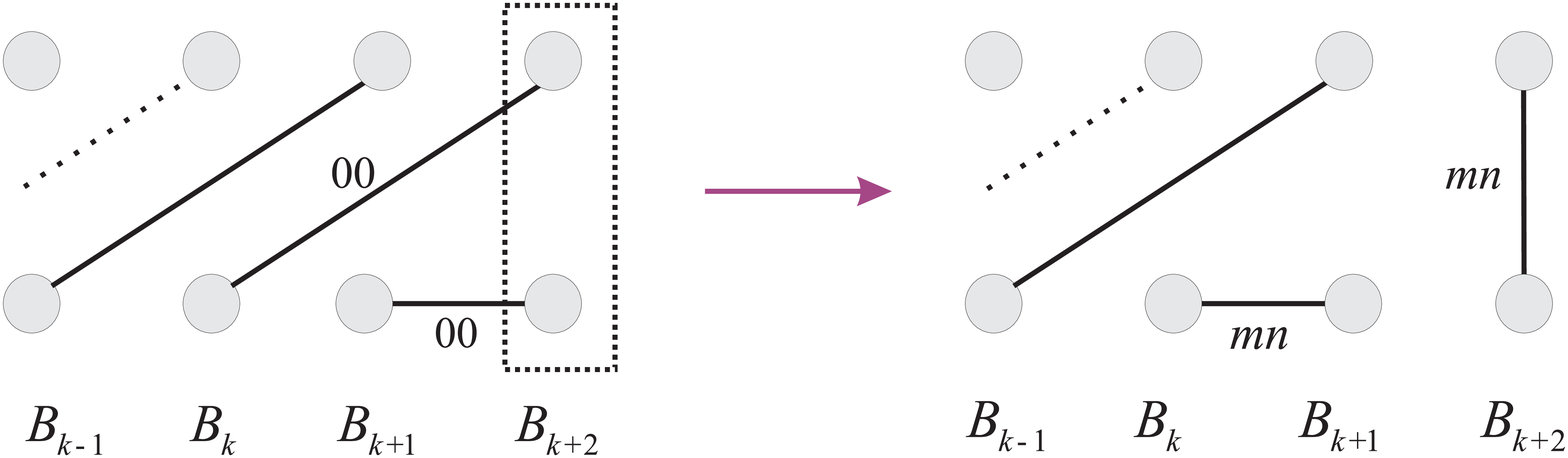}
 \caption{The exchange of the receiver and any of the
 controllers can be achieved by
 entanglement swapping. Here $B_{k+1}$ and $B_{k+2}$ exchange their role, $B_{k+2}$ becomes a controller and $B_{k+1}$ becomes the receiver.
 By a sequence of
 entanglement swapping demanded by Alice, any change of pattern can be induced.}
\label{BellPairPatternsChange}
\end{figure}
\section{Security against cheating}
How this protocol is protected against cheating of probable
dishonest members$?$ If we look at figure (\ref{Passing}), we note
hat any two consecutive members of the group, say $B_{k+1}$ and
$B_{k+2}$ in that figure, may conspire to retrieve the state by
their own, hence cutting off the transfer of state down the chain.
To do this, they need to share another independent Bell pair
between themselves and then use the protocol described in section
($3$) to teleport the state to only one of them say $B_{k+2}$. In
this way and by their own collaboration they can definitely
retrieve the state $\Omega_{k+1}\chi$ which otherwise, could have
been transferred down the chain. In this way they have effectively
been able to cut the flow of state by cutting off the chain.
However in order to retrieve the original state $|\chi\ra$, they
need to know the operator $\Omega$ which depends on all the
results of previous controllers and also that of Alice. Therefore
Alice can defer announcement of her results and also demands that
all the controllers announce the results of their measurements
only after the state has passed through all the chain. One may
argue that the two dishonest parties, can proceed as described
above, keep the state $\Omega_{k+1}\chi$ and send a fake qubit
down the line and then wait until all announcements are made, and
then recover the true state. This possibility is not ruled out,
although it requires that the two dishonest members be located in
adjacent positions of the chain which has after all a low
probability. However, by comparing a random subset of the received
state with the ones sent by Alice, she and the legitimate receiver
can easily detect whether or not if there are dishonest members in
the group. Furthermore by entanglement swapping a few times and
hence changing the receiver and the controllers randomly, she can
not only decrease the role of any cheating, but she can also
detect exactly the location of dishonest members.

\section{A comparative account of necessary resources}
In this section we make a comparison between the resources
necessary for our protocol with that of (\cite{chineese}).
Consider the two-qubit state sharing scheme of \cite{chineese}
with $N$ members ( 1 receiver and N-1 controllers). Figure
(\ref{chineesefigureN}), shows the necessary resources. It
requires 2N Bell pairs and two measurements by Alice in the $N+1$
qubit GHZ basis. Furthermore to announce the results of Alice's
measurements, she needs to announce $2(N+1)$ classical bits and
the
controllers require to announce in total $2(N-1)$ bits. \\

In contrast, as explained in the text and figures
(\ref{BellPairPattern}, \ref{Feeding}), our scheme requires, for the
same number of members, the following resources: First it requires
$N+1$ Bell pairs. Furthermore it requires no measurement in the GHZ
basis. Alice and the controllers require to announce $2(N+1)$
classical bits. Table (\ref{comparison}) summarizes this comparison.

\begin{table}\label{comparison}
\centering
\begin{tabular}{ccccc}
\hline &\vline& The scheme of \cite{chineese} & \vline & our scheme \\

\hline Number of Bell pairs &\vline& 2N & \vline& N+1 \ \    \\
GHZ measurements & \vline& 2  & \vline& 0 \\
Number of classical bits announced by Alice &\vline&  2(N+1) &\vline& 4   \\
Number of classical bits announced by the controllers&\vline& 2(N-1) & \vline& 2(N-1) \\
\hline
\end{tabular}
\caption{Comparison of the resources necessary for two-qubit state
sharing between $N$ members.}
\end{table}

\section{conclusion}
In this article, we have introduced a more efficient and secure
protocol for quantum state sharing of a two-qubit state between a
group of N members. All the members of the group and the sender
should collaborate with each other so that only one member called
the receiver can retrieve the state after receiving classical bits
from the sender and the controllers. The basic ingredient of the
scheme is what we call Passing, where the state is passed through a
chain of Bell pairs hold by the controllers, none of whom can detect
the identity of the state, but whose collaboration is necessary for
the final retrieval of the state by the final receiver. Passing of
the state along the chain is achieved by a sequence of entanglement
swapping. We also discuss the security of the protocol against
cheating of the members of the group and  make a comparison of the
resources necessary for this state sharing with the one given in
\cite{chineese}, which is summarized in table \ref{comparison}. We
hope that the basic idea of state- passing introduced in this paper
can find more applications in other applications and can be
implemented in real experiments.

{}
\end{document}